\documentclass[a4paper]{article}
\usepackage{graphicx}
\usepackage{amsmath}
\usepackage{amsfonts}
\usepackage{amssymb}
\newcommand{\be}{\begin{equation}}
\newcommand{\ee}{\end{equation}}

\newcommand{\al}{\alpha}

\newcommand{\bet}{\beta}

\newcommand{\partialp}{{\cal D}}
\newcommand{\Om}{\Omega}
\newcommand{\om}{\omega}

\newcommand{\G}{\Gamma}

\newcommand{\gh}{\mathfrak}
\newcommand{\eps}{\varepsilon}
\newcommand{\dz}{\wedge}
\newcommand{\C}{{\bf C }}

\newcommand{\ba}{\begin{array}}
\newcommand{\ea}{\end{array}}
\newcommand{\beq}{\begin{eqnarray}}
\newcommand{\eeq}{\end{eqnarray}}
\newtheorem{lm}{Lemma}
\newtheorem{thee}{Theorem}
\newtheorem{proo}{Proposition}
\newtheorem{co}{Corollary}
\newtheorem{rem}{Remark}
\newtheorem{deff}{Definition}
\newcommand{\bd}{\begin{deff}}
\newcommand{\ed}{\end{deff}}

\newcommand{\bl}{\begin{lm}}
\newcommand{\el}{\end{lm}}
\newcommand{\bp}{\begin{proo}}
\newcommand{\ep}{\end{proo}}
\newcommand{\bt}{\begin{thee}}
\newcommand{\et}{\end{thee}}
\newcommand{\bc}{\begin{co}}
\newcommand{\ec}{\end{co}}
\newcommand{\brm}{\begin{rem}}
\newcommand{\erm}{\end{rem}}

\newcommand{\der}{{\rm d}}
\begin{document}
\thispagestyle{empty}
\title{{Projective connections associated with second order ODEs} 
\vskip 1.truecm
\author{Ezra T Newman\\
Department of Physics and Astronomy\\
University of Pittsburgh\\
Pittsburgh PA\\
USA\\
newman@pitt.edu\\
\\
\\
Pawel Nurowski\\
Instytut Fizyki Teoretycznej\\
Uniwersytet Warszawski\\
ul. Hoza 69, Warszawa\\
Poland\\
nurowski@fuw.edu.pl}
}

\author{\mbox{}}
\maketitle
\begin{abstract}
We show that every 2nd order ODE defines a 4-parameter
family of projective connections on its 2-dimensional
solution space. In a special case of ODEs, for which a certain point
transformation invariant vanishes, we find that this family of
connections always has a preferred representative. This preferred 
representative turns out to 
be identical to the projective connection described in Cartan's classic paper
{\it Sur les Varietes a Connection
Projective}.
\end{abstract}

\newpage
\noindent

\rm
\section{Introduction}
\noindent
In recent years there has been a return of interest in the two
related classical issues associated with differential equations: (1) the
equivalence problem (under a variety of transformation types) for the
equations and (2) the natural geometric structures induced by the equations
on their solution spaces. The original studies began, among others, with the
work of Lie \cite{Lie} and his student, Tresse \cite{Tre1,Tre2}. This was
soon followed by W\"{u}nschmann's contribution 
\cite{Wun} and reached its peak with the work of
Cartan \cite{Cart} and Chern\cite{Chern}. Cartan devised an extremely
powerful but difficult scheme for the analysis of the equivalence problem
under the three classes of transformation: fiber preserving, point and
contact. Though equivalence relations were established for a variety of
equations and transformation classes, the calculations were extraordinarily
complicated and long, and as a consequence, many problems were only
partially completed. (The modern advent of algebraic computers has allowed
the completion of many of these problems and opened the door to a variety of
new problems \cite{Kam,Koz,New,Nurg}.) Early in these studies - then confined
to general 2nd and 3rd order odes - it was realized that the equations
themselves defined on the (finite dimensional) solution spaces certain
geometric structures. For example, W\"{u}nschmann discovered that a (large)
class of 3rd order ode's define a conformal (Lorentzian) metric on the
3-dimensional solution space. This class was defined by the vanishing of a
certain function of the 3rd order equation. Later, in the context of
Cartan's and Chern's work, this function was understood as a (relative)
invariant of the equation under contact transformations and became known as
the W\"{u}nschmann invariant. (As an aside we mention that in the modern context
of general relativity, this work was generalized to pairs of 2nd order odes
whose solution space is 4-dimensional. The vanishing of a generalized
W\"{u}nschmann invariant for these equations leads to a conformal Lorentzian
metric on the solution space. All four dimensional Lorentzian metrics are
obtainable in this manner\cite{New,Koz}.)\\

\noindent 
Cartan, following Lie and Tresse, using his scheme for the
analysis of 2nd order odes under point transformations, realized\cite{Cart1}
that a large class of 2nd order odes induced a natural projective structure
on their 2-dimensional solution space. This class was defined (analogously
to the 3rd order ode case) by the vanishing of a certain W\"{u}nschmann-like
function of the 2nd order equation.\\

\noindent 
In the present work we return to the problem of the
geometry associated with any 2nd order ode. Without recourse to Cartan's
equivalence technique, we find that any 2nd order ode defines, via the
torsion-free 1st Cartan structure equation, a 4-parameter family
of projective connections on the solution space.\\

\noindent 
In the second section we review the general theory of normal
projective connections on $n$-manifolds from the point of view of Cartan
connections. We also define projective structures as equivalence
classes of certain sets of one-forms on these manifolds.\\

\noindent 
As an example of projective connections, in the third section, we 
consider the geometry associated with a second order ode. 
We find a natural 4-parameter family of projective connections living
on its two dimensional space of solutions. In general these
connections are quite complicated. They are parametrized by the 
solutions of a certain {\it linear} ode of fourth order, 
which is naturally associated with our ode. 
We find that among all the odes $y''=Q(x,y,y')$ there is a large class
for which the associated 4th order ode is {\it homogeneous}. This
class of equations is characterized in terms of the
vanishing af a certain function constructed solely from $Q$ and its
derivatives, which is directly analogous to the W\"{u}nschman function. 
It turns out that the trivial solution of the homegeneous 4th
order ode singles out a preferred connection from the 4-parameter
family. Then this 
class of second order odes together with this preferred connection
turns out to be identical to the class that Cartan obtained from a
study of the equivalence problem. In the last section we discuss the 
relationship between our and Cartan's method of obtaining this class.\\

\noindent
The work described here is part of a larger project, namely the
study of natural geometric structures induced on the finite dimensional
solution spaces of both odes and certain overdetermined pdes. In earlier work
\cite{etal} we saw how all 4-dimensional conformal metrics and Cartan normal
conformal connections were contained in the space of pairs of pdes
satisfying generalized W\"{u}nschmann equations.  (Similar results hold for all
3rd order odes satisfying the W\"{u}nschmann equation.) In the present work we
have extended these results to unique Cartan normal projective connections
associated with 2nd order odes satisfying a W\"{u}nschmann-like equation.\\

\section{Projective connection}
\subsection{Cartan connection}
\noindent
In this subsection we will first define a {\it Cartan connection} and then
specialize it to a Cartan {\it projective} connection (see \cite{Kob} for
more details).\\

\noindent
Consider a structure $(P,H,M,G)$ such that
\begin{itemize}
\item $(P,H,M)$ is the principal fibre bundle, over an $n$-dimensional
  manifold, with a structure Lie group $H$ 
\item $G$ is a Lie group, of dimension dim$G=$dim$P$, for which $H$ is a
closed subgroup.
\end{itemize}
Denote by $B^*$ the fundamental vector field associated with an element
$B$ of the Lie algebra $H'$ of $H$.
Let $\om$ be a $G'$-valued 1-form on $P$ such that
\begin{itemize}\item[]
\begin{itemize}
\item $\om (B^*)=B$ for each $B\in H'$ 
\item $R_b^* \om=b^{-1} \om b$ for each $b\in H$
\item $\om (X)=0$ if and only if the vector field $X$ vanishes
identically on $P$.
\end{itemize}
\end{itemize}
\vspace{.5truecm}
Then $\om$ is called {\it Cartan's connection on} $(P,H,M,G)$.\\

\noindent
The Cartan {\it projective connection} is a Cartan connection for which
$$ 
G={\bf SL}(n+1,{\bf R})/({\rm center}),~~~~~~~~~~~~~~~~~~~~~~~~~~~~~~~~~~~~~~~~~~~
$$
$$
H=\{\begin{pmatrix}
{\bf A}&0\\
A^T&(\det {\bf A})^{-1}
\end{pmatrix},~{\bf A}\in{\bf GL}(n,{\bf R}),~A\in{\bf R}^n\}/({\rm center})
$$ 

\noindent
In next two subsections we present a convenient way of defining a projective connection on a local
trivialization $U\times H$ of the bundle $P$.

\subsection{Normal projective connection on $U\in M$}

\noindent
Here working on the base space $M$ we define a {\it normal projective
  connection} on $U\subset M$.\\

\noindent
Consider a coframe $(\omega^i)$, $i=1,2,...,n$ on an open
neighbourhood $U$ of $M$. Suppose that in addition you have $n^2$
1-forms $\om^i_{~j}$, $i,j=1,2,...,n$ on $M$ such that 
\be
\der\omega^i+\om^i_{~j}\dz\omega^j=0,~~~\forall i=1,2,...n.\label{tors}
\ee 

\noindent
Then, the system of forms $(\omega^i,\om^i_{~j})$ defines a torsion-free
connection on $U$.

\noindent
Take $n$ {\it arbitrary} 1-forms $(\omega_i)$, $i=1,2,...n$ on
$U$. The forms $(\omega^i,\om^i_{~j},\omega_j)$ define the $n^2$ 2-forms
$\Omega^i_{~j}$ and $n$ 2-forms $\Psi_j$ on $U$ by
\be
\Omega^i_{~j}=\der\om^i_{~j}+\om^i_{~k}\dz\om^k_{~j}+\omega^i\dz\omega_j+\delta^i_{~j}\omega^k\dz\omega_k,
\ee
\be
\Psi_i=\der\omega_i+\omega_k\dz\om^k_{~j}.
\ee 

\noindent
Decompose $\Omega^i_{~j}$ onto the basis $(\omega^i)$,
$$\Omega^i_{~j}=\frac{1}{2}\Omega^i_{~jkl}\omega^k\dz\omega^l.$$ 

\noindent
Find all $(\omega_i)$ for which the so called {\it normal}
condition
\be
\Omega^i_{~jil}=0,~~~~~\forall j,l=1,2,...n \label{normal}
\ee
is satisfied. It turns out that if $n\geq 2$ the forms $\omega_i$ are determined
{\it uniquely} by the equations (\ref{normal}). Indeed, by using the Riemann 2-forms
\be
R^i_{~j}=\frac{1}{2}R^i_{~jkl}\omega^k\dz\omega^l=\der\om^i_{~j}+\om^i_{~k}\dz\om^k_{~j}
\ee
and the Ricci tensor
\be
R_{jl}=R^i_{jil}
\ee
of the connection $\om^i_{~j}$ one finds that
\be
\omega_i=~[~\frac{1}{1-n}R_{(ij)}-\frac{1}{1+n}R_{[ij]}~]~\omega^j.
\ee

\vspace{.5truecm}
\noindent
Having determined the forms $\omega_i$, collect the system of 1-forms
$(\omega^i,\om^i_{~j},\omega_j)$ into a matrix 
\be
\om_u=\begin{pmatrix}
\om^i_{~k}-\frac{1}{n+1}\om^l_{~l}\delta^i_{~k}&\omega^i\\~&~\\
\omega_k&-\frac{1}{n+1}\om^l_{~l}
\end{pmatrix}.
\ee 
Note that $\om_u$ is a 1-form on $U$ which has values in the Lie
algebra $G'={\bf SL}'(n+1,{\bf R})$. It is called 
a {\it normal projective connection} on $U$. 

\subsection{Normal projective connection on $U\times H$}
\noindent
Earlier we defined a Cartan projective connection on the principal
$H$-bundle $(P,M,H,G)$. Here we show how the normal projective
connection on $U\subset M$ can be lifted to $(P,M,H,G)$.\\

\noindent
Choose a generic element of $H$ in the form
\be
b=\begin{pmatrix}
A^i_{~k}&0\\~&~\\
A_k&a^{-1}
\end{pmatrix},
\ee 
where $(A^i_{~j})$ is a real-valued
$n\times n$ matrix with nonvanishing determinant $a=\det (A^i_{~j})$, and $(A_i)$ is a
real row $n$-vector. 

\noindent
Define a $G'$-valued 1-form $\om$ on $U\times H$ by
\be
\om=b^{-1}\om_u b+b^{-1}\der b.\label{konproj}
\ee

\noindent
The 1-form $\om$ defines a {\it projective connection} on $U\times H$. 
This projective connection
on $U\times H$ is called the {\it normal} projective connection. The
term {\it normal} referes to the condition (\ref{normal}), which this
connection satisfies.

\vspace{.5truecm}
\noindent
The explicit formulae for the normal projective connection
(\ref{konproj}) are written below.
\be
\om=\begin{pmatrix}
\om'^i_{~k}-\frac{1}{n+1}\om'^l_{~l}\delta^i_{~k}&\omega'^i\\~&~\\
\omega'_k&-\frac{1}{n+1}\om'^l_{~l}
\end{pmatrix},\label{kontrans}
\ee 
where
\be
\omega'^i=a^{-1}A^{-1i}_{~~~~j}\omega^j,\label{kontransi}
\ee
\be
\om'^i_{~j}=A^{-1i}_{~~~~k}\om^k_{~l}A^l_{~j}+A^{-1i}_{~~~~k}\omega^k
A_j+\delta^i_{~j}A_l A^{-1l}_{~~~~k}\omega^k+A^{-1i}_{~~~~k}\der
A^k_{~j}+\delta^i_{~j}a^{-1}\der a,
\ee
\be
\omega'_i=a ~(~\omega_k A^k_{~i}-A_l A^{-1l}_{~~~~j}\om^j_{~k}A^k_{~i}-A_l
A^{-1l}_{~~~~j}\omega^j A_i+\der A_i-A_l A^{-1l}_{~~~~j}\der A^j_{~i}~)~,\label{kontransf}
\ee
and we have used the fact that 
\be
\der a=a A^{-1l}_{~~~~k}\der A^k_{~l}.
\ee

\vspace{.5truecm}
\noindent
The curvature 
\be
\Om=\der \om+\om\dz\om
\ee
of $\om$ has the form
\be
\Om=b^{-1}\Om_u b,~~~{\rm where}~~~\Om_u=\der\om_u+\om_u\dz\om_u=\begin{pmatrix}
\Om^i_{~j}-\frac{1}{n+1}\delta^i_{~j}\Om^l_{~l}&0\\~&~\\
\Psi_j&-\frac{1}{n+1}\Om^l_{~l}\end{pmatrix}
\ee

\vspace{.3truecm}
\noindent
It is worthwhile to note that if $n\geq 3$ then the vanishing of $\Omega^i_{~j}$ implies
the vanishing of $\Psi_i$. This follows from the Bianchi identity
$\der\Om-\Om\dz\om+\om\dz\Om=0$. 
It is known that in dimension $n=2$ the forms $\Omega^i_{~j}$
are identically equal to zero. In this dimension all the information
about the curvature of the normal projective connection is encoded in the forms $\Psi_i$. 

\vspace{.5truecm}
\noindent
{\bf Remark}\\
To globalize the local trivialization construction of the normal projective
connection described above one needs assumptions about topology of
$M$. In the local treatment we use in this paper these assumptions are not neccessary. 
\vspace{.5truecm}

\subsection{Projective structure on $M$}

\noindent
An alternative view of the formulae
(\ref{kontransi})-(\ref{kontransf}) is to consider them as an
equivalence class of connections on $U$. This motivates the following definition.

\vspace{.5truecm}
\noindent
A {\it projective structure} on an $n$-dimensional manifold $M$ is an
equivalence class $[(\omega^i,\om^i_{~j})]$ of  sets of 1-forms $(\omega^i,\om^i_{~j})$ on $M$ such that
\begin{itemize}
\item $(\omega^i)$, $i=1,2,...,n$ is a coframe on $M$ such that 
$$\der\omega^i+\om^i_{~j}\dz\omega^j=0,~~~\forall i=1,2,...,n$$
\item two sets $(\omega^i,\om^i_{~j})$ and $(\omega'^i,\om'^i_{~j})$
  are in the same equivalence class iff there exists functions
  $A^i_{~j}$ and $A_i$ on $M$ such that 
\be
\omega'^i=a^{-1}A^{-1i}_{~~~~j}\omega^j\label{prst1}
\ee
and 
\be
\om'^i_{~j}=A^{-1i}_{~~~~k}\om^k_{~l}A^l_{~j}+A^{-1i}_{~~~~k}\omega^k
A_j+\delta^i_{~j}A_l A^{-1l}_{~~~~k}\omega^k+A^{-1i}_{~~~~k}\der
A^k_{~j}+\delta^i_{~j}a^{-1}\der a,\label{prst2}
\ee 
with $a=\det (A^i_{~j})\neq 0$ at every point of $M$. 
\end{itemize}

\noindent
It turns out that all the torsion-free
connections from the equivalence class of a given projective structure
have the same set of geodesics on $M$. To see this consider a representative $(\omega^i,\om^i_{~j})$ of a
projective structure on $M$. Let $(e_i)$ be the set of $n$-vector fields dual to the coframe
$(\omega^i)$, i.e. $\omega^i(e_j)=\delta^i_{~j}$. Let $\gamma(t)$ be a geodesic curve for the
connection 1-forms $\om^i_{~j}=\om^i_{~jk}\omega^k$. This means that if
$V=\frac{\der}{\der t}=V^i e_i$ is a
vector tangent to this curve then 
\be
\frac{\der V^i}{\der t}+\om^i_{~jk}V^jV^k=f V^i,\label{geod}
\ee
with a certain function $f$ on $M$. If
$(\omega'^i,\om'^i_{~j})$ belongs to the same projective structure as 
$(\omega^i,\om^i_{~j})$ then the equation
(\ref{geod}) for $V^i$ and the relations between $(\omega^i,\om^i_{~j})$
and $(\omega'^i,\om'^i_{~j})$ imply that in the coframe $(\omega'^i)$ the
$V'^i$ component of the vector $V=V'^i e'_i$ satisfies geodesic equation 
\be
\frac{\der V'^i}{\der t}+\om'^i_{~jk}V'^jV'^k=f' V'^i,\label{geod'}
\ee  
with merely new function $f'=f+2a A_jV'^j$. Thus the curve $\gamma(t)$ is also
a geodesic in connection $\om'^i_{~j}$.
\vspace{.3truecm}

\noindent
Note that if $A^i_{~j}=\delta^i_{~j}$ then 
\be
\omega'^i=\omega^i\label{fixA1}
\ee
and
\be 
\om'^i_{~j}=\om^i_{~j}+\omega^i
A_j+\delta^i_{~j}A,\label{fixA}
\ee 
with $A=A_i\omega^i$. Thus, for a given projective structure
$(\omega^i,\om^i_{~j})$, fixing the coframe does not fix 
the gauge in the choice of $\om'^i_{~j}$. There exists an entire
class (\ref{fixA}) of connections 
that, together with the fixed coframe $(\omega^i)$, represents the
same projective structure. 

\subsection{Equivalence of projective structures}
We say that two projective structures $(\omega^i,\om^i_{~j})$ and
$(\bar{\omega}^i,\bar{\om}^i_{~j})$ on two respective $n$-dimensional
manifolds $M$ and $\bar{M}$ are (locally) {\it equivalent} iff there
exists a (local) diffeomorphism $\phi: M\to \bar{M}$ and functions
$A^i_{~j}$ and $A_j$ on $M$ such that
$$\phi^*(\bar{\omega}^i)=a^{-1}A^{-1i}_{~~~~j}\omega^j$$
and 
$$\phi^*(\bar{\om})^i_{~j}=A^{-1i}_{~~~~k}\om^k_{~l}A^l_{~j}+A^{-1i}_{~~~~k}\omega^k
A_j+\delta^i_{~j}A_l A^{-1l}_{~~~~k}\omega^k+A^{-1i}_{~~~~k}\der
A^k_{~j}+\delta^i_{~j}a^{-1}\der a,$$ 
with $a=\det (A^i_{~j})\neq 0$. \\

\noindent
If, given a projective structure $(\omega^i,\om^i_{~j})$ on
$M$, we have a diffeomorphism $\phi: M\to M$ with $A^i_{~j}$
and $A_j$ as above, such that  
\be
\phi^*(\omega^i)=a^{-1}A^{-1i}_{~~~~j}\omega^j\label{sym1}
\ee
and 
\be 
\phi^*(\om)^i_{~j}=A^{-1i}_{~~~~k}\om^k_{~l}A^l_{~j}+A^{-1i}_{~~~~k}\omega^k
A_j+\delta^i_{~j}A_l A^{-1l}_{~~~~k}\omega^k+A^{-1i}_{~~~~k}\der
A^k_{~j}+\delta^i_{~j}a^{-1}\der a,\label{sym2}
\ee
then we call $\phi$ a symmetry of
$(\omega^i,\om^i_{~j})$. Locally, a 1-parameter group of
symmetries $\phi_t: M\to M$ of $(\omega^i,\om^i_{~j})$ is expressible in terms of the
corresponding vector field $X$, called an {\it infinitesimal
  symmetry}. 
Taking the Lie derivative with respect to $X$ of equations
(\ref{sym1})-(\ref{sym2}) one obtains the following characterization
of infinitesimal symetries. 

\noindent
A vector field $X$ is an infinitesimal symmetry of
a projective structure $(\omega^i,\om^i_{~j})$ iff there
exist functions
$B^i_{~j}$ and $B_j$ on $M$ such that 
\be
{\cal L}_X\omega^i=-(B^i_{~j}+B^k_{~k}\delta^i_{~j})\omega^j\label{symx1}
\ee
\be
{\cal L}_X\om^i_{~j}=\om^i_{~j}B^l_{~j}-B^i_{~l}\om^l_{~j}+\omega^i B_j
+\delta^i_{~j}B_l \omega^l+\der
B^i_{~j}+\delta^i_{~j}\der B^k_{~k}.\label{symx2}
\ee
It is easy to check that a Lie bracket $[X_1,X_2]$ of two
infinitesimal symmetries is an infinitesimal symmetry, hence the
infinitesimal symmetries generate a Lie algebra. This is the Lie
algebra of infinitesimal symmetries of the structure $(\omega^i,\om^i_{~j})$.

\section{Projective structures of second order ODEs}
\subsection{Contact forms associated with a second order ODE}
We now show that a second order ODE defines a projective structure on
the space of its solutions.\\

\noindent
A second order ODE   
\be
\frac{\der^2 y}{\der x^2}~=~Q(~x,~y,~\frac{\der y}{\der x}~)\label{ode}
\ee
for a function ${\bf R}\ni x\to y=y(x)\in{\bf R}$, can be
alternatively written as a system of the two first order ODEs
\be
\frac{\der y}{\der x}~=~p,~~~~~~~~~~~~~~\frac{\der p}{\der x}~=~Q(x,y,p)
\ee 
for two functions ${\bf R}\ni x\to y=y(x)\in{\bf R}$ and  ${\bf R}\ni
x\to p=p(x)\in{\bf R}$. This system defines two (contact) 1-forms 
\be
\omega^1=\der y-p\der x,~~~~~~~~~~~~~~\omega^2=\der p-Q\der x,\label{forms}
\ee 
which live on a 3-dimensional manifold $J^1$, the {\it first jet space}, parametrized by
coordinates $(x,y,p)$. All the information about the ODE (\ref{ode})
is encoded in these two forms. For example any solution  to
(\ref{ode}) is a curve $\gamma(x)=(~x,y(x),p(x)~)\subset J^1$ on
which the forms (\ref{forms}) vanish. 

\vspace{.3truecm}
\noindent
Given an ODE (\ref{ode}), we look for a set
$(\om^1_{~1},\om^1_{~2},\om^2_{~1},\om^2_{~2})$ of 1-forms on $J^1$ such that
\be
\der\omega^1+\om^1_{~1}\dz\omega^1+\om^1_{~2}\dz \omega^2=0,~~~~~~
\der\omega^2+\om^2_{~1}\dz\omega^1+\om^2_{~2}\dz \omega^2=0.\label{torj}
\ee
Introducing the third 1-form 
\be
\omega^3=\der x,
\ee
which together with
$\omega^1$ and $\omega^2$ constitutes a basis of 1-forms on $J^1$, we
find that
\be
\der\omega^1=-\omega^2\dz\omega^3,~~~~~~~\der\omega^2=-(Q_y\omega^1+Q_p\omega^2)\dz\omega^3,
\ee
and that the general solution to the `vanishing torsion' equations
(\ref{torj}) is
\be
\om^1_{~1}=\om^1_{~11}\omega^1+\om^1_{~12}\omega^2,~~~~~~
\om^1_{~2}=\om^1_{~12}\omega^1+\om^1_{~22}\omega^2-\omega^3,\label{fk1}
\ee
\be
\om^2_{~1}=\om^2_{~11}\omega^1+\om^2_{~12}\omega^2-Q_y\omega^3,~~~~~~
\om^2_{~2}=\om^2_{~12}\omega^1+\om^2_{~22}\omega^2-Q_p\omega^3,\label{fk2}
\ee
with some unspecified functions
$(\om^1_{~11},\om^1_{~12},\om^1_{~22},\om^2_{~11},\om^2_{~12},\om^2_{~22})$
on $J^1$. Here, and in the following, we denoted the partial
derivatives with respect to a variable, as a subscript on the function
whose partial derivative is evaluated, e.g. 
$Q_y :=\frac{\partial Q}{\partial y}$. 

\vspace{.3truecm}
\noindent
The anihilator of the contact forms $\omega^1$ and $\omega^2$ is
spanned by the vector field 
\be
D=\partial_x+p\partial_y+Q\partial_p,
\ee
which is defined 
up to a multiplicative
factor. Its integral curves, which coincide with the 
solutions $\gamma(x)$ of the
original equation, are intrinsically
defined. Also the notion of surfaces $S$, {\it transversal}
to $D$ is unambigous.\\

\noindent
Any choice of 1-forms $(\om^1_{~1},\om^1_{~2},\om^2_{~1},\om^2_{~2})$ of
the form given by equations 
(\ref{fk1})-(\ref{fk2}) on the jet space $J^1$ determines projective 
structures $[(\omega^k;\om^i_{~j})_{|S}]$
on each 2-dimensional surface $S$ transversal to $D$. These projective
structures are defined on each
$S$ by transformations (\ref{prst1})-(\ref{prst2}) applied to the 1-forms
$(\omega^k;\om^i_{~j})_{|S}$.
They, in turn, were defined as the restrictions of the 1-forms 
$(\omega^1,\omega^2;\om^1_{~1},\om^1_{~2},\om^2_{~1},\om^2_{~2})$ 
from $J^1$ to $S$. Given a particular choice of functions $\om^i_{~jk}$ in
(\ref{fk1})-(\ref{fk2}) and a pair of transversal to $D$ surfaces $S$ and
$S'$, the projective structures
$[(\omega^k;\om^i_{~j})_{|S}]$ and $[(\omega^k;\om^i_{j})_{|S'}]$ will
be in general inequivalent. It is therefore interesting to ask as to
whether there exist a choice of forms (\ref{fk1})-(\ref{fk2}) which, on
{\it all} transversal surfaces $S$, defines the same (modulo
equivalence) projective structure. Locally, this requirement is
equivalent to the existence of a choice of forms
(\ref{fk1})-(\ref{fk2}) on $J^1$ such that 
the Lie derivative of the forms $(\omega^i;\om^k_{~j})$ along $D$ is
simply the infinitesimal version of the transformations
(\ref{sym1})-(\ref{sym2}). Explicitly, we ask for the existence of
$\om^i_{~jk}$ of (\ref{fk1})-(\ref{fk2}) and the existence of
functions $B^i_{~j}$ and $B_k$ on $J^1$ such that\\
\be
{\cal L}_D\omega^i=-(B^i_{~j}+B^k_{~k}\delta^i_{~j})\omega^j\label{eqons1}
\ee
\be
{\cal L}_D\om^i_{~j}=\om^i_{~j}B^l_{~j}-B^i_{~l}\om^l_{~j}+\omega^i B_j
+\delta^i_{~j}B_l \omega^l+\der
B^i_{~j}+\delta^i_{~j}\der B^k_{~k}~~~~~~i,j=1,2.\label{eqons2}
\ee
If we were able to find a solution $\om^i_{~jk}$ to the above equations,
then it would generate the same projective structure on all
surfaces transversal to $D$. This structure would therefore descend to
the 2-dimensional space of integral lines of $D$ endowing it, or what
is the same, endowing the parameter space of
solutions to the original ODE, with a projective structure.\\



\noindent
To solve equations (\ref{eqons1})-(\ref{eqons2}) we take the most
general forms $(\om^1_{~1},\om^1_{~2},\om^2_{~1},\om^2_{~2})$ (from 
(\ref{fk1})-(\ref{fk2})) that are associated with the ODE. We then use the
gauge freedom (\ref{fixA1})-(\ref{fixA}) preserving   
$$
\omega^1=\der y-p\der x,~~~~~~~~~~~~~~\omega^2=\der p-Q\der x
$$ 
to achieve $$\om^1_{~1}=0$$ everywhere on $J^1$. The forms 
$(\omega^1,\omega^2;\om^1_{~1},\om^1_{~2},\om^2_{~1},\om^2_{~2})$ with
$\om^1_{~1}=0$, when restricted to each $S$, will therefore represent the same
projective structure on $S$ as the original general forms we started with. Thus, without loss of
generality, we solve equations
(\ref{eqons1})-(\ref{eqons2}) for forms 
\be
\omega^1=\der y-p\der x,~~~~~~~~~~~~~~\omega^2=\der p-Q\der x\label{fk20}
\ee
and 
\be
\om^1_{~1}=0,~~~~~~
\om^1_{~2}=\om^1_{~22}\omega^2-\omega^3,\label{fk21}
\ee
$$
\om^2_{~1}=\om^2_{~11}\omega^1+\om^2_{~12}\omega^2-Q_y\omega^3,~~~~~~
\om^2_{~2}=\om^2_{~12}\omega^1+\om^2_{~22}\omega^2-Q_p\omega^3.
$$
It is a matter of straigthforward calculation to achieve the following
proposition.
\bp
The forms (\ref{fk20})-(\ref{fk21}) satisfy equations
(\ref{eqons1})-(\ref{eqons2}) 
{\it if and only if}
\beq
\om^2_{~22}=&&D\om^1_{~22}+2Q_p\om^1_{~22},\nonumber\\
\om^2_{~12}=&&\frac{1}{4}~[-D^2 \om^1_{~22}-3Q_pD\om^1_{~22}+(3Q_y-2Q_p^2-2DQ_p)\om^1_{~22}-Q_{pp}~]\label{sol1}\\
\om^2_{~11}=&&\frac{1}{6}~[D^3\om^1_{~22}+3D^2\om^1_{~22}+(5DQ_p+2Q_p^2-7Q_y)D\om^1_{~22}+\nonumber\\
&&~~~~(2D^2Q_p-3DQ_y+4Q_pDQ_p
  -8Q_pQ_y)\om^1_{~22}+DQ_{pp}-4Q_{py}~]\nonumber
\eeq
and $\om^1_{~22}$ fulfills the differential equation
\be
D^4\om^1_{~22}+a_4 D^3\om^1_{~22}+a_3 D^2\om^1_{~22}+a_2
D\om^1_{~22}+a_1\om^1_{~22}+a_0=0\label{sol2}
\ee
with coefficients $a_0,a_1,a_2,a_3,a_4$ given by
$$a_4=2Q_p,$$
$$a_3=(8DQ_p-Q_p^2-10Q_y),$$
\be
a_2=(7D^2Q_p-10DQ_y+3Q_pDQ_p-2Q_p^3-10Q_pQ_y),\label{sol3}
\ee
$$a_1=
(2D^3Q_p-3D^2Q_y+4(DQ_p)^2+2Q_pD^2Q_p-5Q_pDQ_y-4Q_p^2DQ_p -14Q_yDQ_p
+2Q_p^2Q_y+9Q_y^2),$$
$$a_0=D^2Q_{pp}-4DQ_{py}-Q_pDQ_{pp}+4Q_pQ_{py}-3Q_{pp}Q_y+6Q_{yy}.$$
\ep

\noindent
Thus, modulo equivalence, the only forms (\ref{forms})-(\ref{fk2}) that
generate the same projective structure on all surfaces transversal to
$D$ are given by (\ref{fk21})-(\ref{sol1})
with the coefficient $\om^1_{~22}$ satisfying differential equation
(\ref{sol2})-(\ref{sol3}). Now, recalling that the space of solutions of
the second order ODE can be identified with the 2-dimensional space of
integral lines of $D$ in $J^1$, we obtain the following theorem.

\bt
Every solution $\om^1_{~22}$ to the fourth order
differential equation (\ref{sol2})-(\ref{sol3}) defines a natural
projective structure on the space of solutions $J^1/D$ of the second
order ODE $y''=Q(x,y,y')$. The structure is 
given by the projection from $J^1$ to $J^1/D$ of forms
$$
\omega^1=\der y-p\der x,~~~~~~~~~~~~~~~\omega^2=\der p-Q\der
x
$$
with 
$$
\om^1_{~1}=0,
$$
$$
\om^1_{~2}=\om^1_{~22}\omega^2-\omega^3,~~~~~~~\omega^3=\der x,
$$
\beq
\om^2_{~1}=&&\frac{1}{6}~[D^3\om^1_{~22}+3D^2\om^1_{~22}+(5DQ_p+2Q_p^2-7Q_y)D\om^1_{~22}+\nonumber\\
&&~~~~(2D^2Q_p-3DQ_y+4Q_pDQ_p
  -8Q_pQ_y)\om^1_{~22}+DQ_{pp}-4Q_{py}~]~~\omega^1\nonumber\\
+&&\frac{1}{4}~[-D^2
  \om^1_{~22}-3Q_pD\om^1_{~22}+(3Q_y-2Q_p^2-2DQ_p)\om^1_{~22}-Q_{pp}~]~~\omega^2\nonumber\\
-&&Q_y~~\omega^3,~~~~~~\nonumber
\eeq
\beq
\om^2_{~2}=&&\frac{1}{4}~[-D^2
  \om^1_{~22}-3Q_pD\om^1_{~22}+(3Q_y-2Q_p^2-2DQ_p)\om^1_{~22}
-Q_{pp}~]~~\omega^1\nonumber\\
+&&[~D\om^1_{~22}+2Q_p\om^1_{~22}~]~~\omega^2~~-~~Q_p~~\omega^3.\nonumber
\eeq
\et 

\noindent
Since the equation (\ref{sol2})-(\ref{sol3}) is of 4th order it has four
independent solutions. Thus, all the corresponding
projective structures on $J^1/D$ should be treated on equal
footing. However, in the case of second order ODEs satisfying some
additional conditions, some of these structures may be more
distinguished. In particular, Sophus Lie \cite{Lie} and Elie Cartan
\cite{Cart1} considered
2nd order ODEs satisfying the additional condition
\be
a_0=D^2Q_{pp}-4DQ_{py}-Q_pDQ_{pp}+4Q_pQ_{py}-3Q_{pp}Q_y+6Q_{yy}\equiv 0.
\ee
For such ODEs equation (\ref{sol2})-(\ref{sol3}) is homogeneous and as such
has a preferred solution $\om^1_{~22}=0$. Thus, for this class of
second order ODEs there exists a distinguished, natural projective
structure on $J^1/D$ associated with the solution $\om^1_{~22}=0$ of
(\ref{sol2})-(\ref{sol3}). Explicitely, for any second order ode
satisfying $a_0\equiv 0$, this structure is given by\\
\be
\omega^1=\der y-p\der x,~~~~~~~~~~~~~~\omega^2=\der p-Q\der x\label{lc}
\ee
with 
\be
\om^1_{~1}=0,~~~~~
\om^1_{~2}=-\omega^3,~~~~~\omega^3=\der x,
\ee
\be
\om^2_{~1}=\frac{1}{6}(DQ_{pp}-4Q_{py})\omega^1-\frac{1}{4}~Q_{pp}\omega^2-
Q_y\omega^3,\label{prstrc}
\ee
\be
\om^2_{~2}=-\frac{1}{4}Q_{pp}\omega^1-Q_p\omega^3.\label{lc1}
\ee
\noindent
In general, any projective structure described by Theorem 1 
leads to a projective ${\bf SL}(3,{\bf R})$ connection on an
8-dimensional bundle 
$P\to J^1/D$. One of the features of the projective
structures which via $\om^1_{~22}=0$ are associated with $a_0\equiv 0$, is that each of them
leads to a {\it normal} projective ${\bf SL}(3,{\bf R})$ connection on
$P$. Using the local parameters
$(x,y,p,\al,\beta,\gamma,\nu,\mu)$ for $P$ and equations
(\ref{kontrans})-(\ref{kontransf}), (\ref{lc})-(\ref{lc1}) we find that this ${\bf SL}(3,{\bf
  R})$ connection reads
\be
\om=\begin{pmatrix}
\frac{1}{3}(\Omega_2-2\Omega_1)&-\theta^3&\theta^1\\~&~&~\\
-\Omega_3&\frac{1}{3}(\Omega_1-2\Omega_2)&\theta^2\\~&~&~\\
\Omega_5&-\Omega_4&\frac{1}{3}(\Omega_1+\Omega_2)
\end{pmatrix}\label{mackon}
\ee 
where
$(\theta^1,\theta^2,\theta^3,\Omega_1,\Omega_2,\Omega_3,\Omega_4,\Omega_5)$
are given by
$$
\theta^1=
\al\omega^1,~~~~\theta^2=\bet(\omega^2+
\gamma\omega^1),~~~~\theta^3=\frac{\al}{\bet}(\omega^3+\nu\omega^1),
$$ 
\beq
&&\Omega_1 =\der\log\al-  \mu~\theta^1 + \frac{\nu}{\bet}~\theta^2 - \frac{\bet\gamma}{\al}~\theta^3\nonumber\\ 
&\nonumber
\eeq
\beq
&&\Omega_2=\der\log\bet- \frac{1}{4\al}~[~6\gamma\nu + 4\nu Q_p -
Q_{pp} + 2\al \mu~]~\theta^1 
+ 2\frac{\nu}{\bet}~\theta^2 + \frac{\bet}{\al}~[~\gamma + Q_p~]~\theta^3\nonumber\\
&\nonumber
\eeq
\beq
&&\Omega_3=\frac{\bet}{\al}\der\gamma - \frac{\bet}{6\al^2}~[~DQ_{pp} - 6\gamma^2\nu -
6\gamma\nu Q_p + 3\gamma Q_{pp} - 4 Q_{py} + 6\nu Q_y~]~\theta^1 - \frac{1}{4\al}~[~2\gamma\nu - Q_{pp} + 
2\al \mu~]~\theta^2\nonumber\\
&&- \frac{\bet^2}{\al^2}~[~\gamma^2 + \gamma Q_p - Q_y~]~\theta^3\nonumber\\ 
&\nonumber
\eeq
\beq
&&\Omega_4=\frac{1}{\bet}\der\nu - \frac{1}{6\al\bet}~[~6\gamma\nu^2 + 6\nu^2 Q_p - 3 \nu Q_{pp} +
Q_{ppp}~]~\theta^1 + 
\frac{\nu^2}{\bet^2}~\theta^2 -\frac{1}{4\al}~[~-2\gamma\nu - 4\nu Q_p 
+ Q_{pp} + 2\al \mu~]~\theta^3\nonumber
\eeq
\beq
&&2\Omega_5 = \der \mu+\mu\der\log\al - \frac{\nu}{\al}\der\gamma +
\frac{\gamma}{\al}\der\nu\nonumber\\
&&-\frac{1}{24\al^2}~[~12\al^2 \mu^2 + 48\nu Q_{py} - 48\nu^2 Q_y  - 12\nu
  DQ_{pp} + 36\gamma^2\nu^2 +
48\gamma\nu^2 Q_p - 36\gamma\nu Q_{pp} + 12\gamma Q_{ppp}\nonumber\\
&&+8 DQ_{ppp} + 8 Q_p Q_{ppp} - 12 Q_{ppy} - 
3 Q_{pp}^2~]~\theta^1+ \frac{1}{6\al\bet}~[~6\gamma\nu^2 - 3\nu Q_{pp} + Q_{ppp} + 6\al\nu
\mu~]~\theta^2\nonumber\\
&& - \frac{\bet}{6\al^2}~[~DQ_{pp} - 6\gamma^2\nu 
- 12\gamma\nu Q_p + 3\gamma Q_{pp} - 4Q_{py} + 12\nu Q_y + 6\al\gamma \mu~]~\theta^3.\nonumber
\eeq
The curvature of this connection reads
$$
\Omega=\begin{pmatrix}
0&0&0\\~&~&~\\
0&0&0\\~&~&~\\
\frac{1}{6\al^2\bet}~b_{01}~\theta^1\dz\theta^2&-\frac{1}{6\al\bet^2}~
b_0~\theta^1\dz\theta^2&0
\end{pmatrix},
$$ 
where we have introduced 
$$
b_0=Q_{pppp},~~~~{\rm and}~~~~b_{01}=Db_0 +
(\gamma + 2Q_p) b_0.
$$
The relatively simple form of this curvature agrees with the general theory of normal projective connections
for $n=2$ (compare with the note at the end of section 2.3).\\

\noindent
The next section is devoted to explaining the Lie/Cartan motivation for
considering the class of ODEs leading to the structure defined above.
\subsection{Equivalence classes of 2nd order ODEs modulo point
  transformations}
A point transformation of variables
\be
(~x,~y~)=(~x(\bar{x},\bar{y}),~y(\bar{x},\bar{y})~)\label{point}
\ee
applied to the second order ODE
\be
y''=Q(x,y,y')\label{odenp}
\ee
changes it to the new form 
\be
\bar{y}''=\bar{Q}(\bar{x},\bar{y},\bar{y}').
\ee
The function $Q=Q(x,y,y')$ transforms in a rather complicated way into a new function 
$\bar{Q}=\bar{Q}(\bar{x},\bar{y},\bar{y}')$. But, using appropriate
derivatives of $Q$ one can construct functions which have
nice transformation properties under transformations (\ref{point}). In
particular, the {\it relative invariants}
of the equation (\ref{odenp}) 
are such functions which, under transformations (\ref{point}), scale by a factor. 
Their vanishing is the point
invariant property of the equation. One of such relative invariants is
$$
a_0=D^2Q_{pp}-4DQ_{py}-Q_pDQ_{pp}+4Q_pQ_{py}-3Q_{pp}Q_y+6Q_{yy},  
$$
the same function that appears in equations (\ref{sol3}).
This fact was
already known to Sophus Lie \cite{Lie}. Elie Cartan \cite{Cart1} 
considered the 
problem of finding {\it all} point invariants of (\ref{odenp}). He
used his {\it equivalence method} which, enabled him to determine another
relative invariant
$$
b_0=Q_{pppp}.
$$
Both $a_0$ and $b_0$ are of the same order and, it follows from the
Cartan analysis, that the equation (\ref{odenp}) has no more point
invariants of order less than or equal to 4. Thus, according to Cartan, the second order ODEs modulo point
transformations split into four major classes which are
\begin{itemize}
\item[{\it i})] $a_0=b_0=0$
\item[{\it ii})] $a_0=0$ and $b_0\neq 0$
\item[{\it iii})] $a_0\neq 0$ and $b_0=0$
\item[{\it iv})] $a_0\neq0$ and $b_0\neq 0$.
\end{itemize}
Cases {\it i)-ii)} were analyzed by Cartan completely. In particular, he showed that
if $a_0=0$ then with each point equivalence class of second order ODEs
is associated a natural normal projective connection, whose curvature provides
all the point invariants of the class. This connection equips the
space of solutions of each of the equations from the
equivalence class with a projective structure. It follows that the projective structures
originating in this way from different equations from the same point
equivalence class are equivalent. This distinguished projective
structure associated with the class of equation $y''=Q(x,y,y')$
coincides with the structure (\ref{prstrc}) defined in the previous
section.
\section{Acknowledgments}

\noindent
We acknowledge support from NSF Grant No PHY-0088951 and the Polish
KBN Grant No 2 P03B 12724.

\end{document}